\pdfoutput=1 %

\documentclass{article}

\input{glyphtounicode}
\usepackage{microtype}
\usepackage{graphicx}
\usepackage{subfigure}
\usepackage{booktabs} %

\usepackage{hyperref}
\usepackage{hyperxmp}
\usepackage{accsupp}

\usepackage{xcolor}
\usepackage{colortbl}
\usepackage{lipsum}
\usepackage{xspace}
\usepackage{xurl} %
\usepackage{pifont}
\usepackage{array} %
\usepackage{siunitx} %

\hypersetup{hypertexnames=false}

\ifdefined\toolnamepart\else\def\toolnamepart{full}\fi
\newif\ifabpaper
\newif\ifabappendix
\def\abtmp{paper}\ifx\toolnamepart\abtmp\abpapertrue\fi
\def\abtmp{appendix}\ifx\toolnamepart\abtmp\abappendixtrue\fi
\newcommand{\toolnamecounters}[3]{}
\makeatletter
\newcommand{\toolnamewritecounters}{%
  \immediate\write\@auxout{\string\toolnamecounters
{\the\c@figure}{\the\c@table}{\the\c@lstlisting}}}
\let\ab@Hy@setref@link\Hy@setref@link
\def\Hy@setref@link#1#2#3#4#5#6\@nil#7{%
  \if\relax\detokenize{#4}\relax
  #7{#1}{#2}{#3}{#4}{#5}%
  \else
  \ab@Hy@setref@link{#1}{#2}{#3}{#4}{#5}#6\@nil{#7}%
\fi}
\makeatother
\ExplSyntaxOn
\ior_new:N \g__ab_aux_ior
\cs_new_protected:Npn \__ab_import_labels:n #1
{
  \file_if_exist:nTF { #1 .aux }
  {
    \group_begin:
    \cs_set_protected:Npn \newlabel ##1 ##2
    { \cs_if_exist:cF { r@##1 } { \__ab_newlabel:nnnnnn {##1} ##2 } }
    \cs_set_protected:Npn \toolnamecounters ##1 ##2 ##3
    {
      \bool_if:NT \l__ab_counters_bool
      {
        \int_gset:cn { c@figure } {##1}
        \int_gset:cn { c@table } {##2}
        \int_gset:cn { c@lstlisting } {##3}
      }
    }
    \ior_open:Nn \g__ab_aux_ior { #1 .aux }
    \ior_map_inline:Nn \g__ab_aux_ior
    {
      \tl_if_head_eq_meaning:nNT {##1} \newlabel {##1}
      \tl_if_head_eq_meaning:nNT {##1} \toolnamecounters {##1}
    }
    \ior_close:N \g__ab_aux_ior
    \group_end:
  }
  { \msg_warning:nnn { toolname } { no-full-aux } {#1} }
}
\cs_new_protected:Npn \__ab_newlabel:nnnnnn #1 #2 #3 #4 #5 #6
{ \cs_gset:cpn { r@#1 } { {#2} {#3} {#4} {} {} } }
\bool_new:N \l__ab_counters_bool
\msg_new:nnn { toolname } { no-full-aux }
{ #1.aux~not~found:~build~main.tex~first;~cross-part~references~print~??. }
\AtBeginDocument
{
  \legacy_if:nT { abpaper } { \__ab_import_labels:n { main } }
  \legacy_if:nT { abappendix }
  {
    \bool_set_true:N \l__ab_counters_bool
    \__ab_import_labels:n { main }
  }
}
\ExplSyntaxOff

\IfFileExists{latexml.sty}{\usepackage{latexml}}{\newif\iflatexml}
\usepackage[accepted]{mlsys2025}
\iflatexml
\fi
\makeatletter
\renewcommand{\Notice@String}{Preprint.}
\renewcommand{\printAffiliationsAndNotice}[1]{%
  \stepcounter{@affiliationcounter}%
  {\let\thefootnote\relax\footnotetext{\hspace*{-1.8em}%
      \if\relax\detokenize{#1}\relax\else#1\\\fi
      \forloop{@affilnum}{1}{\value{@affilnum} < \value{@affiliationcounter}}{%
        \ifnum\value{@affilnum}>1 \space\fi
        \textsuperscript{\arabic{@affilnum}}%
      \csname @affilname\the@affilnum\endcsname}\\
      Correspondence: \mlsyscorrespondingauthor@text.\\
\Notice@String}}}
\makeatother

\newcommand{\toolnamekeywords}{GPU-initiated communication, GPU-initiated
  networking, RDMA, InfiniBand, RoCE, GPUDirect Async, IBGDA, NVLink, AWS EFA,
  communication compiler, MLIR, Triton, tile languages, CuTe DSL,
  computation-communication fusion, NVSHMEM, NCCL, NCCL GIN, rocSHMEM,
  all-to-all, Mixture of Experts, MoE, expert parallelism, MoE dispatch and
  combine, DeepEP, vLLM, LLM inference, distributed deep learning, NVIDIA
H100, NVIDIA B200, AMD MI355X}
\hypersetup{
  pdfsubject={Obaleth is a compiler that makes GPU-initiated communication
    part of a kernel's intermediate representation in Triton and MLIR,
    co-optimizes it with computation, and lowers it to specialized RDMA and
  NVLink transport code for NVIDIA and AMD GPUs.},
  pdfkeywords={\toolnamekeywords},
  pdflang={en-US},
  pdfdisplaydoctitle=true,
  keeppdfinfo,
}

\DeclareRobustCommand{\abactual}[1]{%
  \BeginAccSupp{method=pdfstringdef,unicode,ActualText={#1}}%
{\let\toolnamesc\toolnameplain#1}\EndAccSupp{}}
\makeatletter
\let\ab@section\section
\RenewDocumentCommand{\section}{s o m}{%
  \IfBooleanTF{#1}{\ab@section*{\abactual{#3}}}{%
    \IfValueTF{#2}{\ab@section[#2]{\abactual{#3}}}%
{\ab@section[#3]{\abactual{#3}}}}}
\renewenvironment{abstract}
{\centerline{\large\bf\scshape\abactual{Abstract}}
  \vspace{-0.12in}
\begin{quote}}
  {\par
\end{quote}\vskip 0.12in}
\makeatother
\newcommand{\toolnameTitle}{\toolname: Build Your Own GPU Communication Stack}

\mlsystitlerunning{\abactual{\toolnameTitle}}

\usepackage{booktabs,multirow,pifont}

\newcommand{\toolnameplain}{\textsc{Obaleth}}
\DeclareRobustCommand{\toolnamesc}{%
\BeginAccSupp{method=escape,ActualText=Obaleth}\textsc{Obaleth}\EndAccSupp{}}
\newcommand{\toolname}{\texorpdfstring{\toolnamesc}{Obaleth}\xspace}

\newcommand{\cmark}{\ding{51}}
\newcommand{\xmark}{\ding{55}}

\usepackage{listings}

\input{listing-designs}
\UseListingDesign{E}

\input{code-designs}
\UseCodeStyle{2}

\begin{document}

\twocolumn[
  \ifabappendix
  \mlsystitle{\texorpdfstring{\abactual{\toolnameTitle}\\[2pt]%
  \abactual{Appendix}}{\toolnameTitle: Appendix}}
  \else
  \mlsystitle{\texorpdfstring{\abactual{\toolnameTitle}}{\toolnameTitle}}
  \fi

  \mlsyssetsymbol{equal}{*}

  \begin{mlsysauthorlist}
    \mlsysauthor{Javid Baydamirli}{equal,ku}
    \mlsysauthor{Ismayil Ismayilov}{equal,fal}
    \mlsysauthor{Kaan Oktay}{fal}
    \mlsysauthor{Didem Unat}{ku}
  \end{mlsysauthorlist}

  \mlsysaffiliation{ku}{Ko\c{c} University, Turkey}
  \mlsysaffiliation{fal}{fal}

  \mlsyscorrespondingauthor{Javid~Baydamirli}{\mbox{jbaydamirli21@ku.edu.tr}}
  \mlsyscorrespondingauthor{Ismayil~Ismayilov}{\mbox{ismayil@fal.ai}}

  \mlsyskeywords{\toolnamekeywords}

  \vskip \dimexpr 0.3in - 8.4pt\relax

  \ifabappendix\else\begin{abstract}
  Machine learning systems specialize computation to the workload and hardware,
  but communication remains in generic libraries whose device code and transport
  decisions are not visible to high-level compilers. We present \toolname, a
  compiler that makes GPU-initiated communication part of a kernel's
  intermediate representation, co-optimizes it alongside
  computation, and lowers it to specialized transport code. \toolname
  plans completions around buffer lifetimes and optimizes memory ordering,
  synchronization, and batching across NVIDIA and AMD GPUs.
  On H100 GPUs, \toolname sends small messages up to 3.3$\times$ faster,
  and on 256 GPUs runs all-to-all 1.25$\times$ and 2.3$\times$ faster than
  NVSHMEM and NCCL GIN, with much smaller memory and register use. In vLLM, it
  serves Qwen3-235B on 16 B200 GPUs and
  answers requests up to 1.29$\times$ faster than with DeepEP.

\end{abstract}
\fi
]

\printAffiliationsAndNotice{\mlsysEqualContribution} %

\ifabappendix\else
\section{Introduction}
\label{sec:Introduction}

Machine learning systems use specialized compilers to generate kernels for
attention, GEMMs, normalization, and more, just-in-time compiled for tensor
shapes, data types, and the GPU's architecture and
resources~\cite{tvm,pytorch2}. Tile-level
languages are particularly prevalent, from the cross-vendor
Triton~\cite{triton} and TileLang~\cite{tilelang}
to NVIDIA's CuTe DSL~\cite{cute-dsl} and cuTile~\cite{cutile}.

GPU communication, by contrast, lives in libraries of hand-written device
code, such as NVSHMEM~\cite{Nvshmem_Docs} and NCCL~\cite{nccl-228-blog}.
Libraries offer \emph{some} malleability through fine-grained GPU-initiated
functions that can be inlined into computation kernels, but tile DSLs either
link them as LLVM
bitcode, as in Triton-distributed~\cite{triton-distributed} and PyTorch
symmetric memory~\cite{torch-symm-mem}, or call them from generated CUDA
source, as in MPK~\cite{mpk} and DaCe's multi-GPU
backend~\cite{dace-autonomous}. Either way, communication enters at the
final lowering stage, without the DSL's semantic information and outside its
optimization passes.

\begin{figure}[t]
  \centering
  \includegraphics[width=\columnwidth]{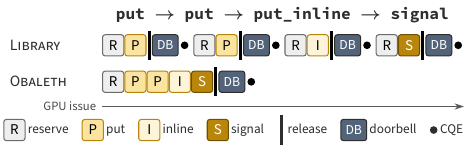}
  \caption{Posting a chain of commands. While a library pays a reservation,
    release, doorbell, and completion per call, \toolname posts the chain
  with one of each.}
  \label{fig:chain}
\end{figure}

\begin{figure}[t]
  \centering
  \includegraphics[width=\columnwidth]{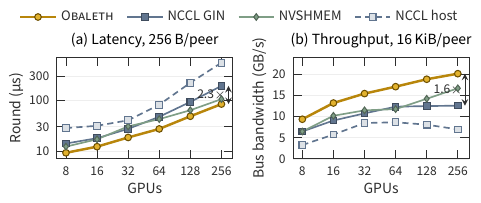}
  \caption{\toolname versus NCCL and NVSHMEM: all-to-all on H100 GPUs.
  }
  \label{fig:teaser}
\end{figure}

Library device functions are written for the general case: they support any
caller, thread geometry, and transport, often through run-time branches, and
must reserve queue slots with atomics and order memory and track completions
conservatively (Figure~\ref{fig:chain}). While each step individually
incurs small costs, together they accumulate: a minimal GPU path issues
an 8-byte put in 0.7\,\si{\micro\second}, and libraries add up to
4.6\,\si{\micro\second} of issue time~\cite{dissecting-gda}.

Higher up the stack,
modern workloads like Mixture of Experts (MoE)
and KV-cache transfer exhibit distinct communication patterns that
cannot be easily expressed with generic point-to-point or collective
calls. Expert parallelism (EP) for MoE alone has produced several competing
libraries, including DeepEP~\cite{deepep2025},
pplx-kernels~\cite{pplx-kernels}, fabric-lib~\cite{pplx-garden-fabric-lib},
UCCL-EP~\cite{uccl-ep}, NCCL EP~\cite{nccl-ep}, Hybrid-EP~\cite{hybrid-ep},
and MORI~\cite{mori}. Similarly, KV-cache transfer has motivated
NIXL~\cite{nixl} and Mooncake~\cite{mooncake}.
Each of these re-implements and tunes parts of the
transport by hand, from work-queue
construction to signaling, completions, and buffer reuse, for one
communication pattern, a set of network interface cards (NICs), and the
shapes it targets.

We present \toolname, a compiler that makes GPU-initiated
communication part of the kernel's intermediate representation (IR),
where workload information is visible.
\toolname represents communication as \op{put},
\op{signal}, and \op{wait} operations in the same IR as the
computation and lowers them progressively into ordinary device code that
builds work-queue entries, rings doorbells, and polls completions.
Instead of packaging each call for the general case, \toolname
specializes to the workload, its inputs, and the network topology,
and generates only the code the transport and NIC need, optimizing
it alongside the computation (Figure~\ref{fig:chain}). We implement
\toolname as MLIR dialects and passes in a plugin for Triton.
Our contributions are as follows:

\begin{itemize}
  \item \textbf{RDMA as IR.} We define layered IRs, including Queue IR, a
    backend-agnostic schedule of submissions and completions, that represent
    GPU-initiated communication
    and lower it to mlx5, AWS EFA, host-proxy, and peer-memory
    backends with GPU and CPU submitters, on NVIDIA and AMD GPUs.
  \item \textbf{Communication optimizations.} We show that program structure
    in the kernel IR enables optimizations that libraries cannot
    perform. \toolname coalesces independent submissions and reorders
    their commands so that a batch pays for one NIC ordering point,
    ties completion and publication to buffer lifetimes, and omits
    atomics and memory ordering when safe. %
  \item \textbf{Tiered adoption model.} We offer two tiers of adoption: a
    generated drop-in device library for existing CUDA and CuTe DSL
    kernels, and custom Triton kernels that fuse communication with
    computation, trading integration effort for optimization scope.
    Adapters describe their communication patterns to the
    compiler.
\end{itemize}

\toolname is faster than communication libraries
and uses less memory. On H100 GPUs, it sends 8-byte messages at up to
3.3$\times$ NVSHMEM's rate and 4.2$\times$ NCCL GIN's. On 256 GPUs, its
small-message all-to-all round is 1.25$\times$ faster
than NVSHMEM and 2.3$\times$ faster than NCCL GIN (Figure~\ref{fig:teaser}).
It also allocates only
22\,MiB of device memory per rank where the libraries allocate
266--584\,MiB. Serving Qwen3-235B~\cite{qwen3} in vLLM~\cite{vllm} on
16 B200 GPUs, it lowers time per output token by 10--24\% relative to
DeepEP and answers requests
1.15--1.29$\times$ faster.
\toolname also has lower latency than the NVSHMEM-based communicators of
Triton and CuTe DSL (1.3--1.5$\times$ in round trip). \toolname is
\href{https://github.com/ParCoreLab/aboleth}{open source}.

\section{Background}
\label{sec:Background}

\paragraph{GPU-initiated communication.}
Kernels can communicate and synchronize without returning control to the
host~\cite{Landscape_Of_GPU_Comm}, sending data as soon as it is produced,
which suits latency-critical and high-message-rate workloads.
Within NVLink domains, kernels use loads and stores on peer-mapped pointers.
Across nodes, remote direct memory access (RDMA)
is the default transport, over InfiniBand, RoCE, or AWS EFA hardware.
GPUDirect RDMA~\cite{gpudirect-rdma} lets NICs access GPU memory directly,
and GPUDirect Async~\cite{GPUDirect-Async,NVIDIA-Blog-IBGDA} lets kernels
trigger network operations themselves.

\paragraph{GPU-initiated RDMA.}
\label{sec:Background-RDMA}
RDMA NICs expose queue pairs
(QPs) of send and receive work queues. To issue an operation such as an RDMA
\code{WRITE}, the submitter writes a work-queue entry (WQE) into the QP's
send queue (SQ).
The shape of this WQE depends on the platform: on NVIDIA ConnectX (mlx5)
hardware, a WQE carries either a data segment
pointing to registered memory or a payload inlined within the
WQE~\cite{Mellanox-PRM-2020}.
The submitter then publishes
the new producer index in a doorbell record (\code{dbrec}) and rings the
doorbell at a NIC register in a user access region (UAR) that is
memory-mapped over PCIe. Once announced, the NIC fetches the WQEs,
reads the payload through DMA (or uses the inlined payload, if
available), performs the operations, and, if requested, writes a
completion-queue entry (CQE) that the submitter then polls.

Other NICs follow the same flow with some differences
in queue placement, ordering, and completion semantics.
AWS EFA~\cite{efa-srd} keeps the SQ in NIC memory that submitters
must write WQEs into with memory-mapped I/O (MMIO) stores, unlike in
mlx5 where the NIC fetches them. Its Scalable Reliable Datagram (SRD)
transport is unordered, unlike mlx5's reliable-connected (RC) queues,
and each of its completions retires only its own request.

\paragraph{Ordering.}
Correct submission requires WQE and payload stores to be visible to the NIC
before the doorbell that announces them and, when one thread rings for WQEs
that others wrote, to the ringing thread first. A system-scope fence that
enforces this costs $3.7\times$ the issue time of a device-scope
one~\cite{dissecting-gda}.

\paragraph{MLIR and tile languages.}
MLIR~\cite{mlir} lets compilers lower progressively through coexisting IRs
(dialects) while reusing analysis and code generation. Triton, CuTe
DSL, and cuTile build on it with dialects that operate on tiles of values.
Triton starts from its target-independent IR (TTIR), lowers it to GPU
IR with explicit layouts (TTGIR), then to LLVM IR, and finally to PTX
(NVIDIA) or AMDGCN (AMD).
\toolname operates at the TTIR stage.

\section{Motivation \& Related Work}
\label{sec:Motivation}

\subsection{Specializing Communication}
NVSHMEM
and rocSHMEM
implement
the OpenSHMEM model, in which ranks allocate a symmetric heap and kernels
issue one-sided operations. NCCL's device API
(GIN)~\cite{nccl-228-blog,gpuinitiatednetworkingnccl}, and
RCCL~\cite{AMD-RCCL-Docs} instead offer one-sided operations on registered
memory windows. Both are written to be compatible with arbitrary
callers, which costs them in the code a kernel carries and in what
each call does at run time. Libraries with several transports either
define separate APIs, as NCCL does,
or keep every transport's code in the kernel,
where it consumes registers even when dormant~\cite{dissecting-gda}. NVSHMEM
offers a build mode without remote transports for this
reason~\cite{nvshmem-p2p-only}.
Environment variables and build options tune a library, but only within the
implementation it ships.

\subsection{Communication in Existing Tile Languages}
Several works enable communication in fused kernels and tile-based
DSLs (Table~\ref{tab:related}). Within peer-accessible memory,
Iris~\cite{iris}, TileLink~\cite{tilelink}, and Pallas Mosaic
GPU~\cite{jax-mosaic-gpu} express transfers as remote loads and
stores, Tensor Memory Accelerator (TMA) copies, or copy-engine operations.
ParallelKittens~\cite{parallelkittens}, FLUX~\cite{flux}, and
Comet~\cite{comet} write them by hand into fused kernels. These systems
leave \emph{some} room for compiler optimizations, but are limited to
a single node. Across nodes, Triton-distributed~\cite{triton-distributed}
and PyTorch symmetric memory~\cite{torch-symm-mem} link NVSHMEM or rocSHMEM
bitcode into Triton kernels and call their device functions. While
Triton-distributed's dialect models
synchronization and symmetric addressing, network operations
remain opaque calls whose issue, ordering, batching,
and completion the compiler cannot change.
Inlining a library's bitcode does not close this gap, as the inlined body
still carries run-time transport dispatch and handling for arbitrary callers.
Compilers that represent
communication explicitly, such as CoCoNet~\cite{coconet} and
Mercury~\cite{mercury}, schedule and fuse collectives but leave the
transport to NCCL. Only on TPUs does Pallas expose remote
DMA beyond peer memory in the IR~\cite{pallas-tpu-distributed}.
To our knowledge, no
compiler represents GPU-initiated network transport
in its IR.

\begin{table}[t]
  \centering
  \caption{GPU communication methods. \emph{In IR}: in a compiler IR.
  \emph{Specialized} by a compiler (\cmark) or by hand (H). $\sim$: partially.}
  \label{tab:related}
  \footnotesize
  \setlength{\tabcolsep}{3pt}
  \begin{tabular}{@{}>{\raggedright\arraybackslash}p{1.3cm}lccc@{}}
    \toprule
    & System & In IR & Specialized & Inter-node \\
    \midrule
    \multirow{3}{=}{Libraries}
    & *SHMEM   & \xmark & \xmark & \cmark \\
    & NCCL/RCCL GIN      & \xmark & \xmark & \cmark \\
    & PyTorch SymmMem    & $\sim$ & $\sim$ & \cmark \\
    \cmidrule{1-5}
    \multirow{3}{=}{Workload-specific}
    & DeepEP             & \xmark & H      & \cmark \\
    & FLUX               & \xmark & H      & $\sim$ \\
    & Comet              & \xmark & H      & \xmark \\
    \cmidrule{1-5}
    \multirow{4}{=}{Tile-based}
    & ParallelKittens    & \xmark & H      & \xmark \\
    & Iris               & $\sim$ & \cmark & \xmark \\
    & TileLink           & $\sim$ & $\sim$ & \xmark \\
    & Triton-distributed & $\sim$ & $\sim$ & \cmark \\
    \cmidrule{1-5}
    \multirow{3}{=}{Compilers}
    & Mosaic GPU         & \cmark & \cmark & \xmark \\
    & CoCoNet            & \cmark & $\sim$ & \cmark \\
    & Mercury            & \cmark & \xmark & \cmark \\
    \midrule
    \multicolumn{2}{@{}l}{\textbf{\toolname}} & \cmark & \cmark & \cmark \\
    \bottomrule
  \end{tabular}
\end{table}

\subsection{Workload-Specific Communication Libraries}
Generic communication serves communication-heavy workloads poorly,
forcing developers to write and tune their own transport
code~\cite{megascale-moe}. In MoE EP, each layer
\emph{dispatches} every token to the ranks that host its top-$k$
experts and \emph{combines} the expert outputs back, creating a
sparse, data-dependent pattern. A plain all-to-all must pad every
peer's buffer to a fixed capacity~\cite{gshard,switch} or first
exchange counts with a host synchronization~\cite{pplx-blog}, permute
tokens into per-peer buffers~\cite{nccl-ep}, and send a token once per
expert unless it is deduplicated~\cite{deepseekv3,uccl-ep}.
DeepEP~\cite{deepep2025} illustrates the effort that specialized
libraries invest instead. Its low-latency kernels map one RC queue
to each local expert, and its high-throughput kernels deduplicate
tokens per node and forward them over NVLink. Other EP libraries make
different choices for portability,
submission path, and topology.  No single library wins everywhere,
and each one has an optimal configuration of token counts, hidden
sizes, expert counts, network topology, and hardware.
Disaggregated inference~\cite{distserve,splitwise} repeats the story for
KV-cache transfer~\cite{nixl,mooncake}, and reinforcement-learning
post-training for weight
synchronization~\cite{pplx-garden-fabric-lib,tensorhub}.

\paragraph{Hand-written GPU RDMA does not port.}
\label{sec:Fragile}
Hand-written GPU RDMA code is tied to a NIC family and a deployment. On AMD
GPUs, MORI and rocSHMEM each maintain separate device code for NVIDIA
ConnectX, Broadcom Thor 2, and AMD Pollara
NICs~\cite{Broadcom_Thor_2,AMD_Pollara_400,AMD-rocSHMEM-Docs,mori}. Even on
the same hardware, the submission
path depends on the deployment environment: the GPU can ring
doorbells only if the NIC's doorbell register is mapped into its address
space, which the NVIDIA kernel module restricts to root unless module options
permit it. Without it, NVSHMEM falls back to a host thread that forwards
doorbells~\cite{Nvshmem_Docs}, and other libraries delegate submission to
CPU proxy threads~\cite{gpuinitiatednetworkingnccl,uccl-ep}.

\subsection{The Opportunity}
\label{sec:Opportunity}
Consider a fused MoE dispatch kernel that quantizes token blocks, sends them
to the ranks hosting their experts, and then sends each peer a token count
and a ready flag.
With a device library, every write in this kernel is a
complete posting sequence (Figure~\ref{fig:chain}): an atomic
reservation, a wait for earlier reservations, a fence, a doorbell, and later
a completion poll. Libraries pay for each step on every call because they
cannot see the following facts in the kernel IR:
(i)~\textbf{What follows the call}: the chunks, count, and
flag sent to a peer form one chain, chains to different experts are
independent, and together they can share one reservation, release, doorbell,
completion, and NIC ordering point, with the count inside the WQE.
(ii)~\textbf{Who else issues}: one program sends to each peer, so reserving
slots needs no atomic and announcing them requires no wait for other producers.
(iii)~\textbf{How long buffers live}: a staging slot is overwritten only every
$P$ launches, so the sender needs a completion only that often.
These facts are not specific to one workload, and they guide the
optimizations of \toolname.

\section{\toolname Overview}
\label{sec:Overview}

\begin{figure}[t]
  \centering
  \includegraphics{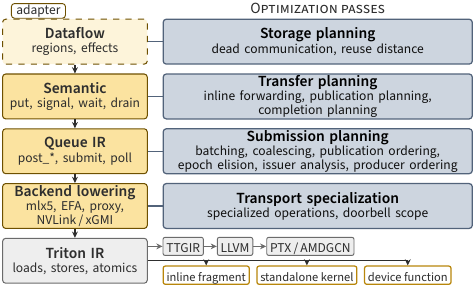}
  \caption{\toolname compilation pipeline.
  Optimization passes rewrite each IR level using the facts it exposes.}
  \label{fig:arch}
\end{figure}

\toolname compiles semantic GPU communication into transport code specialized
alongside the rest of the kernel and decides how transfers are submitted,
published, and completed. Figure~\ref{fig:arch} shows its compilation pipeline.

\subsection{Programming Model}

\paragraph{Endpoints and regions.}
Communication operations act on \emph{regions} and travel over
\emph{endpoints}. An endpoint is a stream of operations from one
peer to another and holds the submission and completion state. It is backed
by a transport queue --- an RC queue pair on mlx5, an SRD queue pair
on EFA, a proxy ring, or a peer-memory mapping. Operations on
different endpoints are unordered, and a rank may be served by
several independent endpoints.
\emph{Regions} are local or remote buffers carrying the address and keys
the transport needs.
Regions are ordinary allocations and need not be symmetric.

\begin{table}[t]
  \centering
  \footnotesize
  \caption{Core operations in \code{obaleth.language} (\code{e}: endpoint,
    \code{t}: token). Transfers also take a \code{predicate}, transfers and
  signals an \code{after} token, and \op{wait} may declare buffer reuse.}
  \label{tab:design-ops}
  \begin{tabular*}{\columnwidth}{@{}l@{\extracolsep{\fill}}l@{}}
    \toprule
    \textbf{Operation} & \textbf{Description} \\
    \midrule
    \op{endpoint(table, i)} & Endpoint $i$ of a table \\
    \op{put(e, dst, src, n)} & Write $n$ bytes from \code{src} \\
    \op{put_inline(e, dst, x)} & Write register value $x$ \\
    \op{signal(e, flag, v)} & Publish epoch $v$ to \code{flag} \\
    \op{wait(t)} & Sender completion of \code{t} \\
    \op{wait_signal(flag, v)} & Wait until \code{flag} $\geq v$ \\
    \op{drain(e)} & Complete all work on \code{e} \\
    \bottomrule
  \end{tabular*}
\end{table}

\paragraph{Operations.}
Table~\ref{tab:design-ops} lists \toolname's core operations. \op{put} and
\op{put_inline} are used to transfer data, the latter embedding small
payloads in the WQE where supported, which saves the NIC a DMA
read and needs no sender completion. A \op{signal} publishes an epoch
that tells the
receiver that data has arrived or that a buffer is free. Signals may
join a completion group (\op{begin_group}/\op{wait_group}) that
completes many submissions with one wait, and a \op{wait} may declare
that its sources are reused only every $P$ launches (\op{wait(t,
iteration=i, reuse=P)}). Transfers can be disabled with a false
\code{predicate} without changing the operation sequence, so data-dependent
communication keeps a static shape that the compiler can batch.

\paragraph{Semantics.}
\toolname operations guarantee:

\begin{itemize}%
  \item \textbf{Ordering.} Transfers and signals return a \code{token}, a
    compiler handle for a communication dependency. Passing it as
    \code{after} orders the new operation after it on the same endpoint.
    Operations not ordered by tokens are unordered.
  \item \textbf{Sender completion.} \op{wait(token)} returns once the
    sources of \code{token} and of every operation ordered before it may
    be overwritten. It makes no guarantees to the receiver.
  \item \textbf{Receiver visibility.} \op{wait_signal(flag, v)} returns
    once the local \code{flag} holds at least \code{v}, and the transfers
    ordered before the signal that wrote \code{v} are guaranteed to be visible.
\end{itemize}

The \code{token} also determines how a signal is published: a signal
ordered after transfers is \emph{strong} and carries their
visibility, while a standalone signal, such as an acknowledgment, is
\emph{weak}.
These guarantees hold on unordered transports such as EFA as well,
and are deliberately minimal, as each leaves the
compiler room to optimize. Since operations that no token orders are
unordered, the
compiler may merge independent chains and reorder their commands
(\S\ref{sec:Opt-Coalescing}), and since \op{wait_signal} waits for
\emph{at least} an epoch, it may skip epochs that a later one covers.

\paragraph{Transports.}
\toolname takes the transport choice as a compile-time input and
specializes the generated code to it. When every peer uses one transport,
e.g., in an NVLink domain, the kernels contain only the code for it.
With mixed transports, e.g., NVLink \emph{and} RDMA,
\toolname partitions the communication into transport-homogeneous
segments and compiles each one directly, so no Tier-2 operation
dispatches on the transport at run time.
The transport is similarly specialized to payload shapes and sizes
through Triton's just-in-time compilation.

\subsection{Adoption Tiers}
\label{sec:Tiers}
\toolname offers two adoption tiers that differ in how much of the
program the compiler sees and can therefore optimize.

\paragraph{Tier-1: Library.}
At this tier, the compiler sees one operation at a time, along with
the peer topology. \toolname generates an NVSHMEM-style device
library of one-sided operations specialized for the selected transports
and GPU. This tier plugs into existing
kernels in CUDA, CuTe DSL, Triton, and other contexts that allow
linking external device code (Appendix Listing~\ref{lst:tier1}).
Isolated calls still contain only the code their transport needs. In mixed
deployments, a call selects the transport by peer, in a branch that folds
away for statically known peers. The library ships as LLVM bitcode, a
relocatable archive, and a generated header.

\paragraph{Tier-2: Custom kernel.}
At this tier, users compose \toolname operations in custom Triton kernels. Unlike
at Tier-1, the compiler can now see communication together with the
computation that produces and uses it. This includes constant sizes, pointer
offsets, predicates, and transfer dependencies.
Listing~\ref{lst:tier2} shows a fused FP8 MoE dispatch
kernel. The kernel contains information beyond simple communication
operations, such as a static loop, data movement and transformations,
and a quantization step, all of which are visible to \toolname.
From this, the compiler derives that the chunk transfers, the count,
and the flag form one submission, that the staged count can be inlined
in the work request, and that one thread owns each queue
(\S\ref{sec:Passes}).

\begin{lstlisting}[style=pystyle, float=t, caption={Tier-2: a fused FP8 MoE dispatch kernel in Triton.\listingnote}, label={lst:tier2}]
@triton.jit
def dispatch(x, counts, endpoints, send, recv, recv_n,
             flag, epoch, RANK, BLOCK, CHUNK, CHUNKS,
             SLOTS):
  peer = tl.program_id(0)    # RANK..SLOTS: tl.constexpr
  ep = obaleth.endpoint(endpoints, peer)  # one issuer
  dst = obaleth.remote(recv, peer)
  n = tl.load(counts + peer)          # tokens for peer
  slot, t = epoch % SLOTS, None
  for c in tl.static_range(CHUNKS):   # unrolled
    q, s = quantize_fp8(load_rows(x, peer, c, n))
    src = send.at(offset(slot, peer, c))
    store_chunk(src, q, s)            # staging
    t = obaleth.put(ep, dst.at(offset(0, RANK, c)), src,
          CHUNK, predicate=c * BLOCK < n, after=t)
  cnt = send.at(count_offset(slot, peer))
  store_count(cnt, n)                 # staged count
  t = obaleth.put(ep, obaleth.remote(recv_n, peer)
          .at(RANK), cnt, 4, after=t) # -> inline
  s = obaleth.signal(ep, obaleth.remote(flag, peer)
          .at(RANK), epoch, after=t)
  obaleth.wait(s, iteration=epoch, reuse=SLOTS)
\end{lstlisting}

While the portion of the code fed to the compiler uses Triton as the frontend,
the generated code is not confined to it. \toolname exports a compiled
communication program in three forms: an \textbf{inline fragment} that can be
used in another Triton
kernel, fusing the communication into its computation; a
\textbf{device function}
with bitcode, relocatable objects, and a generated header, which CUDA kernels
call like a Tier-1 operation; or a \textbf{standalone kernel} with
launch metadata that can be launched in a stream.

\section{Compiler Implementation}
\label{sec:Implementation}

\toolname consists of MLIR dialects and passes, a Python frontend with C++
bindings, and a runtime that creates and manages the endpoints and regions.
The dialects and passes are loaded into Triton as a third-party plugin.
The runtime bootstraps the transport with an all-gather of rank topology,
queue-pair addresses, and region keys, provided by \code{torch.distributed}
or a TCP store.

\subsection{RDMA as IR}
\label{sec:IR}

\toolname IR represents communication at three levels inside a kernel, plus a
dataflow level for framework patterns (Figure~\ref{fig:arch}).
Semantic operations (Table~\ref{tab:design-ops}) act on endpoints, regions,
and tokens that the frontend creates from Triton pointers and keys.
They are lowered to Queue IR, which defines the schedule of the
communication, and then to Triton IR, from which the hardware code is generated.

\paragraph{Queue IR.}
Queue IR is a backend-agnostic schedule of the commands that each endpoint
executes. It defines \code{post_put}, \code{post_inline}, and \code{post_signal}
that construct ordered commands, which a \code{submit} then announces.
A single submission includes one queue reservation, one or more chains of
commands (\emph{segments}), each closed by its own signal, one announcement
(i.e., a doorbell ring), and at most one completion. Commands within a
segment execute in the order determined by the tokens, and submissions on one
endpoint are announced in order.
Attributes on \code{submit} record the passes' plans
(Appendix~\ref{sec:App-Opts}, Listing~\ref{lst:queue-ir}).

Having a separate schedule IR has two benefits. First, transport schedules are
optimized independently of each target's descriptor format, ordering rules,
and completion model. Passes query the transport's capabilities ---
ordered delivery, remote atomics, inline payloads, aggregate completions ---
and switch their optimizations accordingly. $N$ optimizations on $M$
backends therefore do not need $N \times M$ implementations. Second, most
optimizations are independent of the frontend and reachable through adapters.

\paragraph{Framework adapters.}
Integrations whose compute kernels live outside Triton, such as an inference
engine's MoE layer, can describe their computation and communication
as a \emph{pattern}: the logical regions, the transfers among them,
and the region effects of the compute phases. At the dataflow level
of the IR (Figure~\ref{fig:arch}), passes remove unneeded transfers
and staging copies, retarget outputs into final storage, and compute
buffer reuse (Appendix~\ref{sec:Opt-Adapters}). The transfers become
Tier-2 kernels that pass through the usual pipeline, and the
framework launches them.

\subsection{Triton Integration}
\label{sec:Triton}
\toolname is an out-of-tree Triton plugin whose module of builtins,
\code{obaleth.language}, emits \toolname operations (\op{obaleth.put},
\dots) into TTIR. \toolname's passes run after
Triton's target-independent optimizations (inlining,
common-subexpression elimination (CSE), loop unrolling) and
before conversion to TTGIR, where GPU layouts are assigned.
Communication operations are scalar
in Triton IR: one thread of the program publishes each submission, and Tier-1
functions issue from each calling thread. Triton then
lays out the communication code, allocates registers, and compiles it to the
GPU's assembly together with the computation. The transport and planning
decisions join Triton's cache key, so each specialization is compiled
as a separate kernel.

\subsection{Optimizations}
\label{sec:Passes}

\toolname's optimizations rewrite the semantic operations and Queue IR. A round
of communication costs time on both the GPU that issues it and on the
NIC that executes it, and the passes below target both.
We measure the contribution of each pass in \S\ref{sec:Eval-Ablation}.
The complete list of passes is provided in Appendix Table~\ref{tab:impl-opts}.

\begin{figure}[t]
  \centering
  \includegraphics[width=\columnwidth]{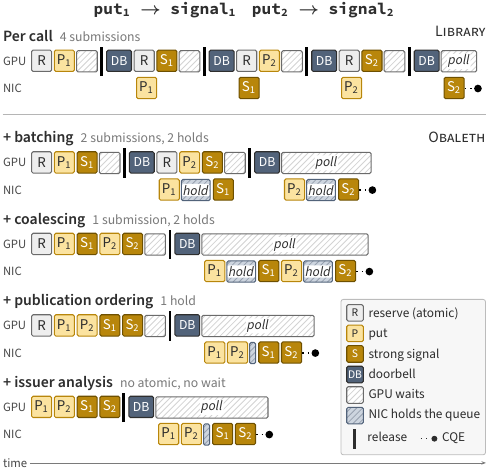}
  \caption{\toolname's optimization passes, described in
    \S\ref{sec:Opt-Coalescing}, applied cumulatively to two
    independent chains on one
    endpoint, removing per-call submissions, NIC
  holds, and atomics.}
  \label{fig:batching}
\end{figure}

\paragraph{Submission batching.}
Batch formation walks each signal's \code{after} chain back through the
transfers it depends on and replaces the chain with one \code{submit} at the
signal and one \code{poll} at the matching \op{wait}. Computation and the
stores that produce later chunks may sit between the transfers, as in
Listing~\ref{lst:tier2}, as long as the chain stays in one block and on one
endpoint with nothing blocking or synchronizing in between. Disabled
transfers stay in the batch as no-op commands. On mlx5, a chain of $N$
commands thus costs one reservation, $N$ contiguous WQEs, one doorbell, and
at most one CQE, instead of $N$ complete posting sequences.

\paragraph{Submission coalescing.}
\label{sec:Opt-Coalescing}
Kernels often issue several independent chains on one endpoint, e.g., a
dispatch program sends tokens and a flag to each expert hosted on its peer.
Coalescing merges such submissions into one: the chains share one
reservation, release, doorbell, and completion, while each keeping its own
signal (Figure~\ref{fig:batching}). Earlier chains are then announced
later, which is legal because the
chains are mutually unordered.
With 16 chains per round, coalescing cuts the GPU's issue
time from 41.5 to 3.5\,\si{\micro\second}, but the total time drops
only from 43.8 to 39.8\,\si{\micro\second}, as the bottleneck
instead shifts into the NIC.

\paragraph{Publication ordering.}
Coalescing exposes the NIC-side cost of strong signals. Each signal
is an ordering point: on mlx5, the NIC holds
the queue at the signal until the payload ahead of it has landed. We measure
this cost at about 2.3\,\si{\micro\second} for both remote atomics and
fenced writes.
A merged batch laid out chain-by-chain ($P_1 S_1
\ldots P_K S_K$) thus stalls $K$ times. Publication ordering rewrites such
chains to have all payloads precede the signals
(Figure~\ref{fig:batching}). With this, every signal still follows
its own payload,
but the submission only stalls once.
The optimization takes the 16-chain
round time from 39.8 to 4.3\,\si{\micro\second}.
The pass needs ordered queues; EFA instead retires the payload before each
strong signal (\S\ref{sec:Backend-Lowering}).

While libraries also batch, they do so only when asked: NCCL GIN
aggregates requests when the caller flags every put but the
last~\cite{nccl-2297-notes,nccl-gin-flags}, and even then each
put-with-signal places its ordering point behind its own payload.

\paragraph{Issuer analysis.}
When several threads may post to one queue, each must reserve its slots with
an atomic and, after writing its WQEs, wait until every earlier reservation
has been written before it rings the doorbell.
A library cannot rule this
case out, so every call pays for it.
\toolname posts each submission from one thread of its program, so it
only needs to know whether two programs can select the same endpoint.
Issuer analysis checks the index passed to each \op{obaleth.endpoint}.
If all selections from a table use the same index, and that index is a
one-to-one function of the program ID (Appendix~\ref{sec:App-Backends}),
each endpoint has a single issuer. The reservation then becomes a plain
add, and the wait is removed.
NCCL offers the same fast path as an exclusive context that the programmer
requests~\cite{nccl-2297-notes}, while \toolname derives it from the IR.
Without the analysis, the one-chain stream is 1.6$\times$ slower and the
ping-pong 1.7\,\si{\micro\second} slower per round trip.

\paragraph{Completion planning.}
\label{sec:Opt-Lifetimes}
Completions ensure that a source buffer may be reused and that a queue slot
may be overwritten. Libraries complete conservatively whenever the caller
asks: NVSHMEM's \code{quiet} polls every configured queue, used or not,
which can double its latency~\cite{dissecting-gda}. \toolname instead defers
completions to where they are needed. In Listing~\ref{lst:tier2},
\code{reuse=SLOTS} declares how many iterations apart its sources are
reused, and completion
planning turns the \op{wait} into a poll every $k$-th iteration
(Appendix Figure~\ref{fig:completion}), where
$k$ is the largest power of two within the reuse distance, the number of
iterations whose commands fit in the queue, and the transport's preferred
interval (16 on mlx5 and EFA). The submission requests a CQE only
on the last command, which, on mlx5, retires the commands before it.
On EFA, the poll counts completions.

\paragraph{Epoch elision.}
Each iteration still has one strong signal and thus one NIC ordering point.
In a loop that completes every $k$ iterations, epoch elision publishes
signals only on completing iterations and the last, turning others into
no-ops of the same size. This is legal because \op{wait_signal} waits for
\emph{at least} an epoch, and a strong signal also makes earlier payloads
visible. As receivers may see an epoch up to $k-1$ iterations late, the pass
skips kernels that wait on a peer.

\subsection{Backend Lowering}
\label{sec:Backend-Lowering}

Backends lower Queue IR to Triton loads, stores, atomics, arithmetic, and
structured control flow, with inline PTX/AMDGCN where Triton lacks the required
memory ordering. Descriptor formats, opcodes, ordering scopes, and completion
formats are fixed at compile time; only queue addresses are read from the
endpoint descriptor at run time (Appendix~\ref{sec:App-Backends},
Table~\ref{tab:backends}). The queues the GPU builds for mlx5
are announced either \emph{GPU-direct}, with the kernel writing the doorbell,
or \emph{CPU-assisted}, through a host-pinned cell that a host thread
forwards (\S\ref{sec:Fragile}). The runtime probes this once per endpoint.
The generated code is identical for both submitters, as only the doorbell
address changes.

\paragraph{mlx5.}
The emitter generates the RDMA flow described in
\S\ref{sec:Background-RDMA} directly in Triton IR. As every command's
size is a compile-time constant, each WQE sits at a constant offset
in its reservation, even inline writes that span several 64-byte WQE blocks.
A strong signal is published as a remote compare-and-swap,
which the responder cannot pass writes ahead of
~\cite{Mellanox-PRM-2020}, even when the payload is registered with
relaxed ordering. It is a fenced write if every receiver registration is
strictly ordered.
A weak signal is a plain
write.

\paragraph{EFA.}
SRD transport is reliable but unordered and has no remote
atomics~\cite{efa-srd}, so a signal could overtake its payload. To keep
strong signals strong, \toolname retires the payload
before it posts the epoch, which costs an extra round trip and bounds how
far sender completion can be deferred.
Our AMD backend, which uses no remote
atomics either, publishes the same way (\S\ref{sec:Eval-AMD}).

\paragraph{Peer memory.}
Over peer memory like NVLink or CUDA IPC, a \op{put} becomes a copy of
8-byte loads and stores over all warps
(Appendix~\ref{sec:App-Backends}), an inline payload a store from registers,
and a signal a system-scope release exchange on the peer's flag. When a
signal follows the copy in the same submission, its release also publishes
the copied tile, so the copy ends with a block barrier instead of a system
fence.

\section{Evaluation}
\label{sec:Evaluation}

\begin{figure*}[t]
  \centering
  \includegraphics[width=\textwidth]{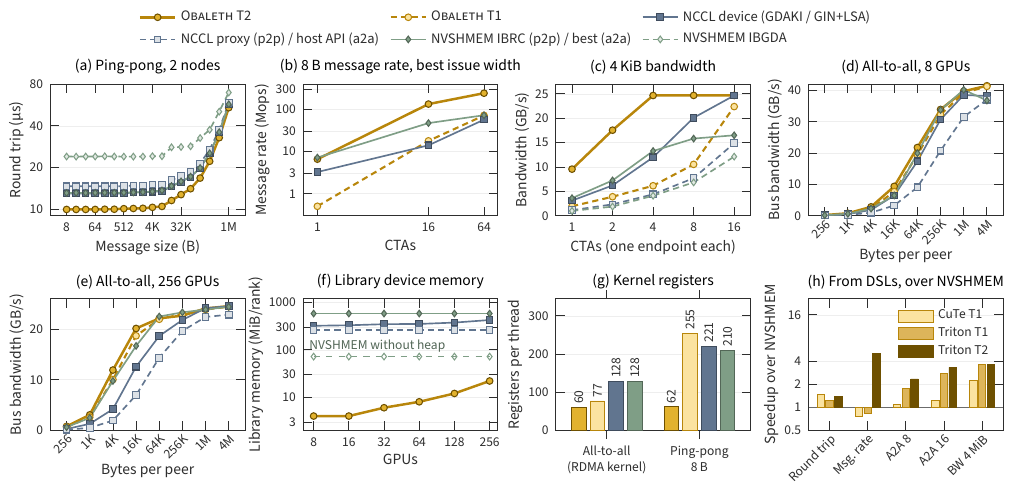}
  \caption{Comparison with NCCL and NVSHMEM on H100 GPUs. (a--c)
    Point-to-point between two nodes: (b) each library at its best issue width and
    transport, (c) one issuing thread per CTA. (d, e) All-to-all bus
    bandwidth on 8 and 256 GPUs (4 GPUs per node).
    (f) Least device memory per rank of any configuration while connecting,
    without the benchmark's buffers but with NVSHMEM's symmetric heap
    (dashed: without it). (g) Registers per
    thread of the all-to-all's RDMA kernel and the ping-pong kernel.
    (h) \toolname from CuTe DSL and Triton over each DSL's NVSHMEM
    (\S\ref{sec:Eval-DSL}). Tier-1 uses a GPU-scope doorbell in (b) at 16 and
    64 CTAs, (d), and (h), and a system-scope one elsewhere; (a) shows Tier-2
    only.
  }
  \label{fig:libraries}
\end{figure*}

\subsection{Setup}
\label{sec:Eval-Setup}

\paragraph{Platforms.}
Appendix Table~\ref{tab:platforms} lists hardware and software
versions in detail. The H100 platform has four GPUs per node,
connected by NVLink within
a node and by 200\,Gb/s InfiniBand across nodes. This platform does not let
GPUs ring doorbells, so \toolname and the libraries
build queues and WQEs on the GPU and forward only the doorbell
through a host thread (CPU-assisted, \S\ref{sec:Backend-Lowering}).
The B200 platform, used for the vLLM experiments and to confirm the
library comparison with GPU doorbells,
has eight GPUs per node, 400\,Gb/s ConnectX-7 InfiniBand, and the
GPU rings the doorbells (GPU-direct, \S\ref{sec:Backend-Lowering}).

\paragraph{Baselines.}
We compare against NCCL's host and device APIs (NCCL GIN, with
GPU-side GDAKI and proxy backends), as well as NVSHMEM's InfiniBand
GPUDirect Async (IBGDA) and IBRC transports. Each library runs its best
configuration for each
experiment, e.g., fused put-with-signal and warp-cooperative
operations, or a packaged collective. \toolname runs as Tier-1 from
CUDA and as Tier-2 Triton kernels. The EP experiments compare against
DeepEP~\cite{deepep2025} 2.5.0 in its default configuration, in the same
vLLM build. On H100, we use NVSHMEM 3.4.5, as 3.8.0 is
2.1\,\si{\micro\second} slower per round trip there.

\paragraph{Methodology.}
In-kernel timings use \code{\%globaltimer},
and host-side APIs use CUDA events. Unless noted, each point is the
median of five interleaved runs of at least 1000 iterations after 200
warmup iterations, and we verify flags and payloads after timing.
For the MoE experiments, both backends serve the same model and
expert kernels, with CUDA graphs and dual-batch overlap. Requests
arrive at a fixed rate and decode to their natural end of sequence.
Before any timing, each model's outputs are validated against an
independent FP32 reference.

\subsection{Comparison with NCCL and NVSHMEM}
\label{sec:Eval-Libraries}

\toolname has lower latency, a higher message rate, and, up to 16\,KiB per
peer, a higher all-to-all throughput than both libraries, the last also with
GPU doorbells, with a fraction of their memory and registers.

\paragraph{Point-to-point.}
Figure~\ref{fig:libraries}a shows the point-to-point latency between
two nodes with one GPU each. Tier-2 has the lowest round trip at
every size, about 3\,\si{\micro\second} ahead of both libraries:
9.98\,\si{\micro\second} at 8\,B, against
12.74 for NCCL GIN and 13.12 for NVSHMEM on H100 GPUs. Most of the
gap comes from the reservation
atomic and system-scope doorbell fence that \toolname omits
(\S\ref{sec:Eval-Ablation}).

The benefit of batching shows most in the message rate
(Figure~\ref{fig:libraries}b). With
each library at its best issue width, Tier-2 sends 8-byte messages on 16
thread blocks (CTAs), each with its own queue, at 2.8$\times$ NVSHMEM's
rate, and on 64 CTAs at 243\,M\,msg/s, 3.3$\times$ NVSHMEM's rate and
4.2$\times$ NCCL GIN's, because it rings one doorbell per window where the
libraries ring one per put.
It also reaches the line rate at 4\,KiB with 4 CTAs, where NCCL GIN needs 16
(Figure~\ref{fig:libraries}c). The libraries lead only where there
is little to batch or complete: on a single CTA, and in a pure stream without
completions, where NVSHMEM posts from 128--256 threads per queue and Tier-2
from one (Appendix~\ref{sec:App-Eval}).

\paragraph{All-to-all.}
Each GPU exchanges with three NVLink peers within its node and over RDMA with
the rest. Up to 16\,KiB per peer, Tier-2 leads in latency and throughput at
every scale. At 256\,B per peer, it is 1.32$\times$ faster than
NVSHMEM's packaged collective and 1.52$\times$ faster than NCCL GIN on 8
GPUs, and 1.25$\times$ and 2.3$\times$ on 256
(Figure~\ref{fig:teaser}),
where its bus bandwidth at 16\,KiB per peer is 1.6$\times$ NCCL GIN's and
1.2$\times$ NVSHMEM's. From 64\,KiB per peer, the exchange
is NIC-bound: the time to move the
payload over the link outweighs the per-message cost,
and the device APIs converge, within 4\% of each other from 1\,MiB
(Figure~\ref{fig:libraries}e).

These leads hold with GPU doorbells: on B200, Tier-2 is 1.11--1.16$\times$
faster than NCCL GIN
and 1.24--1.52$\times$ faster than NVSHMEM at 256\,B on 16 and 32 GPUs
(Appendix Figure~\ref{fig:app-b200}). Part of the larger lead over
NCCL GIN on H100
thus comes from its host-forwarded doorbells. 

\paragraph{Resources.}
\toolname's runtime allocates only 4--22\,MiB of device memory per rank on
8--256 GPUs, at least 12$\times$ less than either library, for its queues and
endpoint descriptors
(Figure~\ref{fig:libraries}f). NCCL allocates 266--426\,MiB for its
communicator and GIN contexts, and NVSHMEM 584\,MiB, of which 512\,MiB
is the symmetric heap it maps on every rank, however little of it the
application uses.
\toolname's RDMA kernels, which carry one transport's code for one call pattern,
also need fewer registers (Figure~\ref{fig:libraries}g). The all-to-all
uses 60 per thread, against 128 in library kernels, and the
ping-pong uses 62~vs.~210--221. Tier-1's generic ping-pong kernel uses 255.

\begin{figure}[t]
  \centering
  \includegraphics[width=\columnwidth]{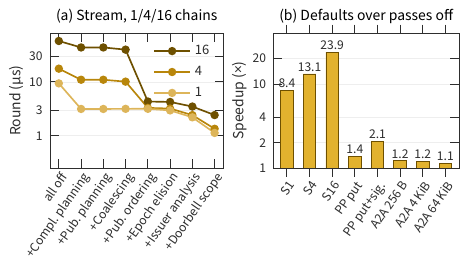}
  \caption{Passes in Tier-2 kernels. (a) Streams of 1, 4, and 16 chains of
    a 64\,B put and a signal, passes added in pipeline order. (b) Defaults
    over all switchable passes off. S$n$: $n$-chain stream; PP: ping-pong
    of eight 64\,B puts, answered by a put or a signal each. A2A:
    All-to-all on 8 GPUs.
  }
  \label{fig:ablation}
\end{figure}

\subsection{Ablation of Compiler Passes}
\label{sec:Eval-Ablation}

We switch each pass on and off in Tier-2 kernels (Figure~\ref{fig:ablation})
and, since one pass's savings can stay hidden behind another's bottleneck,
report both a waterfall that adds passes in pipeline order and the slowdown
when one pass is removed from the defaults (Appendix
Figure~\ref{fig:app-loo}).

On a stream of chains, the defaults are 8.4$\times$ faster than no passes with
one chain and 23.9$\times$ with 16. With one chain, waiting for a completion
every round dominates, and completion planning gives 76\% of the
gain. With 16 chains, the NIC's
ordering points dominate, and coalescing with publication ordering gives 71\%
(removing coalescing, which also disables ordering, costs 5.8$\times$).
Passes also unmask one another: epoch elision's waterfall step is only 7\%
while issuing on the GPU still dominates the round, but once issuer analysis
and the GPU-scope doorbell remove that cost, the NIC's ordering point is the
bottleneck and removing elision costs 2.4$\times$. In the ping-pong, issuer
analysis and the GPU-scope doorbell give the whole 2.1$\times$ gain with a
signal per put, and publication planning saves 0.7\,\si{\micro\second} per
round trip.

In the all-to-all on 8 GPUs, the switchable passes add only
1.13--1.22$\times$, as each peer receives one put and one signal per round
and there is nothing to coalesce or elide. Most of Tier-2's lead comes from
the lowering that is always on: with every switchable pass off, it still
beats NCCL GIN (11.5 against 14.7\,\si{\micro\second} at 256\,B) and Tier-1
(15.9\,\si{\micro\second}), mainly by posting one batched put-and-signal
submission per peer from straight-line code.
The remaining Tier-1 gap is retirement: with one issuer per endpoint and a
GPU-scope doorbell, Tier-1 waits for its \op{quiet} each round, taking
12.1\,\si{\micro\second} to Tier-2's 9.2. Retiring every 4th or
32nd round cuts it to 8.1--9.6\,\si{\micro\second}, and Tier-2 retiring
each round takes 13.5.
A library sees no reuse distance and must retire when asked.

\paragraph{Other platforms.}
\label{sec:Eval-AMD}
The passes act on Queue IR, so they run unchanged on other backends. On AMD
MI355X GPUs with AMD Pollara RoCE NICs, our GPU-direct backend runs
a pipelined stream 3.8--5.7$\times$ faster with the passes than without.
Even through a host proxy, \toolname has the lowest round trip up to
32\,KiB there, and its all-to-all is 1.24--1.70$\times$ and
2.2--2.5$\times$ faster than rocSHMEM's and RCCL's on 8 and 16 GPUs
(Appendix~\ref{sec:App-AMD}). Our EFA backend is also implemented.
We leave full AMD and EFA evaluations to future work.

\subsection{MoE Expert Parallelism in vLLM}
\label{sec:Eval-MoE}

\begin{table}[t]
  \centering
  \footnotesize
  \caption{Qwen3-235B-A22B in vLLM on 16 B200 GPUs (EP16), 128
    GSM8K~\cite{gsm8k} requests per rate. \toolname / DeepEP, medians unless
    noted. Answer: request latency; TTFT: time to first token; TPOT: time
  per output token.}
  \label{tab:moe-qwen}
  \setlength{\tabcolsep}{4pt}
  \begin{tabular}{@{}lrrr@{}}
    \toprule
    \textbf{Requests per second} & \textbf{1} & \textbf{2} & \textbf{4} \\
    \midrule
    Answer (s) & 5.66 / 7.31 & 6.50 / 8.29 & 8.10 / 9.29 \\
    \quad Speedup & 1.29$\times$ & 1.28$\times$ & 1.15$\times$ \\
    TTFT (ms) & 125 / 134 & 127 / 140 & 131 / 134 \\
    TTFT p95 (ms) & 376 / 192 & 312 / 218 & 311 / 201 \\
    TPOT (ms) & 21.9 / 28.7 & 25.9 / 33.2 & 33.4 / 37.3 \\
    Correct (of 128) & 123 / 124 & 124 / 120 & 120 / 122 \\
    \bottomrule
  \end{tabular}
\end{table}

We integrate \toolname into vLLM as an all-to-all backend for expert
parallelism. The integration is a 384-line adapter that
guides the compiler with the application configuration: token bound,
hidden size, top-$k$, expert placement, data type, and how the expert
kernels read their inputs. It calls the compiled dispatch and
combine kernels, whose buffers, shape variants, and queues
are derived by pattern passes
(Appendix~\ref{sec:Opt-Adapters}). Everything
else stays with vLLM.
The adapter states only facts vLLM already knows and, unlike with DeepEP,
no communication kernel is written or tuned by hand for this workload.

On Qwen3-235B-A22B (BF16, EP16), \toolname answers requests
1.15--1.29$\times$ faster than DeepEP (Table~\ref{tab:moe-qwen}). The gain is
largest in decode, where each step's exchange is small and latency-bound, and
the time per output token is 10--24\% lower, matching the kernels'
1.7--3.1$\times$ lead at one token per rank (\S\ref{sec:Eval-DSL}).
The tail of the time to first
token is worse (1.4--2.0$\times$ at p95), and accuracy is similar.
With the FP8 Qwen3-30B-A3B checkpoint on four GPUs of one node, \toolname answers
1.21--1.48$\times$ faster, again with a higher time to first token. On long
GovReport prompts~\cite{govreport}, where prefill dominates, it answers
1.33$\times$ faster.

The gains come from the protocol specialized to the shapes at hand. In decode,
the compiled plan has a capacity of one token, one queue per peer, and four
commands per peer and phase, where DeepEP keeps 128 progress channels and
updates each, empty or not. Over three decode steps of
Ling-1T~\cite{ling} (EP16),
\toolname's NICs therefore transmit 5.6$\times$ fewer packets, and its send and
await kernels take 4.3$\times$ less GPU time than DeepEP's dispatch and
combine, which also fuse layout work.

The pattern passes show the computation guiding the communication. Because
the pattern states how the expert kernels read their inputs, storage planning
sizes each buffer to what they reach, which recovers 16\,GiB per GPU,
45\% more KV-cache capacity than without the pass
and within 5\% of DeepEP's, at no cost in answer time. Shape specialization
compiles each call for the live batch instead of the declared bound, and
valid extents send only live rows, cutting the bytes sent by 58\%. The other
pattern passes change the kernels' service time by less than 3\%.
These results use one EP configuration at rates that saturate neither
engine; it is not the best deployment of these models, as two TP8 replicas
of Qwen3-235B serve long reports with 2.4$\times$ \toolname's throughput.
\toolname's dispatch packs in a separate kernel here; the fused form of
Listing~\ref{lst:tier2} runs at 0.78--0.93$\times$ its speed on 8 H100 GPUs
in BF16 and 0.44--0.82$\times$ with FP8 quantization.

\subsection{Communication from Other Kernel DSLs}
\label{sec:Eval-DSL}

We port our microbenchmarks to CuTe DSL and Triton and compare against the
NVSHMEM each DSL ships with: NVSHMEM4Py (CuTe) and
Triton-distributed~\cite{triton-distributed} (Figure~\ref{fig:libraries}h,
Appendix~\ref{sec:App-DSL}).
Called from a DSL, the generated library keeps its lead: with a GPU-scope
doorbell, Tier-1's round trip
from CuTe and Triton (10.8\,\si{\micro\second}) is lower than NCCL
GIN's and NVSHMEM's even from CUDA in the same runs (12.7 and
13.0\,\si{\micro\second}).
Tier-2 adds batching and leads Triton-distributed, at its faster NVSHMEM
transport for each benchmark, by 1.4$\times$ in round trip,
2.3--3.3$\times$ in all-to-all, and 5.1--10$\times$ in message rate.
\toolname allocates 4\,MiB per rank in every language, against 2.1 and
1.6\,GiB for NVSHMEM4Py and Triton-distributed.

\begin{figure}[t]
  \centering
  \includegraphics[width=\columnwidth]{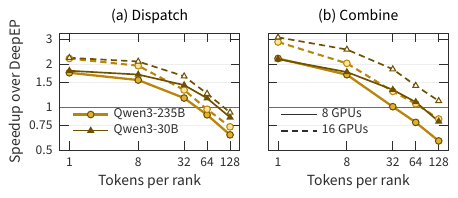}
  \caption{Compiled MoE dispatch and combine over DeepEP on H100, one FP8
  layer, median of five runs (times: Appendix Figure~\ref{fig:app-dsl-moe}).}
  \label{fig:dsl-moe}
  \vspace{-8pt}
\end{figure}

For one MoE layer's FP8 dispatch and combine, the kernels compiled from the
vLLM pattern (\S\ref{sec:Eval-MoE}) beat DeepEP's hand-tuned kernels at the
small batches of
decode: at one token per rank, by 1.7--2.2$\times$ on dispatch and
2.2--3.1$\times$ on combine. From 64 tokens on Qwen3-235B, DeepEP is faster
by up to 1.7$\times$ (Figure~\ref{fig:dsl-moe}).
That cost lies in the compiled kernels, not the transport. Around cuTile
pack and reduce kernels, Tier-2 beats NCCL's \code{all_to_all} by up to
1.87$\times$ in dispatch and 2.43$\times$ in combine, and from 32 tokens this
combine beats the compiled one by up to 1.5$\times$.

\section{Conclusion}
\label{sec:Conclusion}

\toolname brings GPU-initiated communication into the kernel's IR, where it
is co-optimized with the computation. With semantic operations and a
backend-agnostic Queue IR, the compiler batches and reorders submissions,
plans completions around buffer lifetimes, and omits atomics and memory
ordering, independently of NIC hardware. \toolname plugs into CUDA, CuTe DSL,
and Triton kernels, and into frameworks through pattern adapters. It
outperforms vendor libraries with a fraction of their memory, speeds up
end-to-end MoE inference, and runs on NVIDIA and AMD GPUs.
Code is available \href{https://github.com/ParCoreLab/aboleth}{on GitHub}.

\fi

\ifabappendix\else
\section*{Acknowledgements}
\label{Acknowledgements}

Authors from Ko\c{c} University have received funding from the European
Research Council (ERC) under the European Union's Horizon 2020 research and
innovation programme (grant agreement No~949587). We acknowledge the EuroHPC
Joint Undertaking for awarding access to the MareNostrum5 supercomputer in
Spain where the experiments reported in this paper were partially performed.

\fi

\ifabappendix
\appendix

\section{Additional Compiler Optimizations}
\label{sec:App-Opts}

Table~\ref{tab:impl-opts} is the complete list of optimization passes,
each with the cost it removes, requirements, and tiers it applies to.
Listing~\ref{lst:queue-ir} shows the Queue IR that
Listing~\ref{lst:tier2} compiles to.

\begin{figure}[t]
  \centering
  \includegraphics[width=0.85\columnwidth]{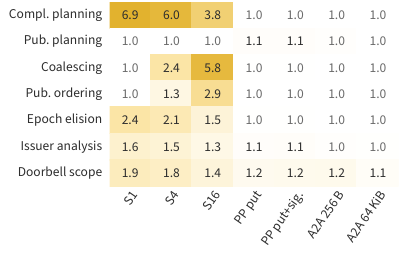}
  \caption{Slowdown of Tier-2's defaults with one pass removed, for the
  workloads of Figure~\ref{fig:ablation}.}
  \label{fig:app-loo}
\end{figure}

\begin{table*}[t]
  \centering
  \footnotesize
  \caption{\toolname's optimization passes. Facts are derived from the IR (D),
    declared by the kernel or framework (K), or reported by the runtime before
    compilation (R). F marks Tier-2 kernels compiled from a framework
  pattern. $^\ast$Off by default.}
  \label{tab:impl-opts}
  \setlength{\tabcolsep}{5pt}
  \begin{tabular}{@{}l>{\raggedright\arraybackslash}p{0.29\textwidth}>{\raggedright\arraybackslash}p{0.31\textwidth}cl@{}}
    \toprule
    \textbf{Optimization} & \textbf{Removes} & \textbf{Requires} &
    \textbf{Tier} & \textbf{Where} \\
    \midrule
    Transport specialization & Run-time transport dispatch, dormant code &
    Peer transports (R) & 1, 2 & --- \\
    Doorbell scope & System-scope release & Queue placement (R) & 1, 2 &
    --- \\
    Publication planning & Ordering point of payload-free signals &
    Tokens (D) & 1, 2 & --- \\
    Inline forwarding & Staging store, source key, NIC DMA read &
    Small constant payload staged by scalar stores (D) & 2 &
    App.~\ref{sec:App-Opts} \\
    \addlinespace[3pt]
    Submission batching & Per-command reservation, doorbell, CQE &
    Token chain (D) & 1, 2 & \S\ref{sec:Opt-Coalescing} \\
    Submission coalescing & Per-chain reservation, doorbell, CQE &
    Independent chains on one endpoint (D) & 2 & \S\ref{sec:Opt-Coalescing} \\
    Publication ordering & Per-signal NIC ordering point &
    Independent chains (D), ordered queue (R) & 2 &
    \S\ref{sec:Opt-Coalescing} \\
    \addlinespace[3pt]
    Issuer analysis & Reservation atomic, wait for other producers &
    Injective endpoint index (D) & 2 & \S\ref{sec:Passes} \\
    Producer ordering & Redundant block barriers &
    Writes since the last barrier (D) & 2 & App.~\ref{sec:App-Opts} \\
    \addlinespace[3pt]
    Completion planning & CQEs and polls between reuses &
    Reuse distance (K), queue depth (R) & 2 & \S\ref{sec:Opt-Lifetimes} \\
    Epoch elision & Ordering points between completions &
    Completion plan, loop (D) & 2 & \S\ref{sec:Opt-Lifetimes} \\
    Queue reclamation & Waits for a poll when the queue is full &
    Kernel-predicated wait (K), single issuer (D) & 2 &
    App.~\ref{sec:App-Opts} \\
    Doorbell deferral$^\ast$ & Doorbells between completions &
    Completion plan, loop (D) & 2 & App.~\ref{sec:App-Opts} \\
    \addlinespace[3pt]
    Shape specialization & Work and traffic padded to the bound &
    Bound range, uniform live bound (K) & F & App.~\ref{sec:Opt-Adapters} \\
    Valid extents & Transfers of dead padding &
    Valid count (K) & 2, F & App.~\ref{sec:Opt-Adapters} \\
    Storage planning & Unreached storage, staging copies &
    Region effects (D), consumer reads (K) & F & App.~\ref{sec:Opt-Adapters} \\
    Dead communication & Transfers that no consumer reads &
    Consumer graph (D) & F & App.~\ref{sec:Opt-Adapters} \\
    Endpoint fan-out & One queue per large chain &
    Byte bound (D), platform profile (R) & F & App.~\ref{sec:Opt-Adapters} \\
    Receive coalescing & One receiver program per lane &
    Lane plan (D) & F & App.~\ref{sec:Opt-Adapters} \\
    \bottomrule
  \end{tabular}
\end{table*}

\begin{lstlisting}[style=cudastyle, float=t, caption={Tier-1: a CUDA kernel calling the generated device library.\listingnote}, label={lst:tier1}]
// world = obaleth.World(rank, size, bootstrap=b)
// world.connect_all("auto")
// regions = world.exchange_regions({...})
// lib = world.device_library(specialize=True)
#include <obaleth/device.cuh>
using namespace obaleth;

__global__ void exchange(DeviceContext ctx, int peer,
    RemoteRegion recv, LocalRegion send, uint32_t n,
    RemoteRegion flag, const uint64_t* mine,
    uint64_t epoch) {
  if (threadIdx.x != 0) return;
  put_signal(ctx, peer, recv, send, n, flag, epoch);
  quiet(ctx, peer);              // source reusable
  wait_signal(ctx, mine, epoch); // data arrived
}
\end{lstlisting}

\begin{lstlisting}[style=mlirstyle, float=t, caption={Queue IR the compiler produces for Listing~\ref{lst:tier2} with \code{CHUNKS = 2} staged token chunks per peer, after batching and planning. Types are omitted.\listingnote}, label={lst:queue-ir}]
%c0 = obaleth.post_put %ep, %dst0, %src0, %sz,
        %live0
%c1 = obaleth.post_put %ep, %dst1, %src1, %sz,
        %live1, %c0
%cn = obaleth.post_inline %ep, %cnt, [%n], %true, %c1
%fl = obaleth.post_signal %ep, %flag, %epoch, %cn
%s  = obaleth.submit %ep, %c0, %c1, %cn, %fl
    {batchSize = 4, exclusiveIssuer, producerBarrier}
obaleth.poll %s, %due {coalesced} // k-th epochs
\end{lstlisting}

\paragraph{Inline forwarding.}
Kernels often stage small values, such as counts and offsets, in a buffer only
to immediately send them. When a \op{put}'s source is written only by scalar
stores just before it, and its size is a constant of at most 28 bytes
(one 64-byte mlx5 WQE block), \toolname deletes the
stores and sends the values from registers. On mlx5, the payload is put
inside the WQE, and over peer memory, it is sent as direct stores. The
transfer then no longer needs a staging store, source key, or sender
completion, and the NIC is spared a DMA read.

\paragraph{Producer ordering.}
When every thread of the block writes the payload and one thread announces
it, the writers' stores must be ordered before that thread's release, a
barrier libraries leave to the caller. \toolname tracks the writes other
threads may have made since the last barrier and places one barrier before
the first submission that reads them. Submissions with no
new writes in between share it, and inline payloads need no ordering.

\begin{figure}[t]
  \centering
  \includegraphics[width=\columnwidth]{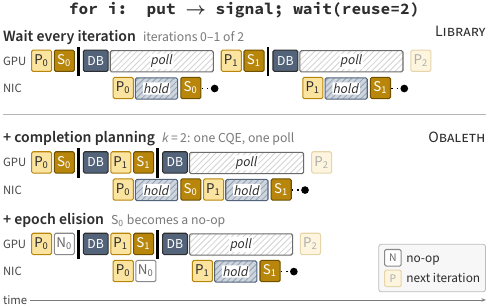}
  \caption{Completion planning and epoch elision on one chain with
  \texttt{wait(reuse=2)}. Symbols as in Figure~\ref{fig:batching}.}
  \label{fig:completion}
\end{figure}

\paragraph{Queue reclamation.}
A kernel may defer completion with a predicate,
\op{wait(t, predicate=p)}, to complete only after its last window, as
NVSHMEM and NCCL callers do when they stream. Nothing then bounds the deferral
by the queue's depth, and a submission that finds the queue full would wait
for a poll that never comes. When one program issues on each endpoint, the
mlx5 lowering lets that submission retire completed submissions itself:
deferred submissions request a CQE whether or not their poll is due, and the
capacity check consumes CQEs in order until the batch fits. Planned
completion is unaffected, as its interval already fits the queue. With
reclamation, a Tier-2 kernel streams
8\,B messages at 213\,M\,msg/s on 16 queues and reaches the NIC's
256\,M\,msg/s on 64 (Appendix~\ref{sec:App-Eval}).

\paragraph{Doorbell deferral.}
Doorbells can be deferred, for example, to the end of a loop. Commands
in iterations between completions are written to the queue but announced
only with the next completing iteration, so the GPU rings once per
completion, rather than once per iteration. This pass improves performance
when issue time dominates, but hurts it when the bottleneck is NIC-bound.
With one chain per round, doorbell deferral reduces the round from
1.12 to 0.81\,\si{\micro\second}, whereas with 16 chains, it slows it down
from 2.41 to 3.06\,\si{\micro\second}.

\subsection{Programs from Framework Patterns}
\label{sec:Opt-Adapters}

Frameworks know facts beyond the kernel IR, such as the maximum number
of tokens a step can carry, what other compute kernels read, and the fact
that a batch's live size is the same on every rank. These can be stated
by patterns and guide the passes in this section. The pattern produces
a plan that Tier-2 kernels are compiled with.

\paragraph{Valid extents.}
A region may declare that only a prefix of run-time length is valid and
that the rest is dead padding with \code{region.valid(count, capacity,
size)}. The compiler then clamps each \op{put} by both its
requested bytes and the valid bytes that remain in its source so that
batches only send live rows. Peer-memory copies are clamped the same way.

\paragraph{Shape specialization.}
From a declared range of bounds and the fact that the live bound is uniform
across ranks, the compiler emits variants at power-of-two capacities and a
selector keyed by the live bound. Each variant's storage, endpoint plan, and
completion interval follow from its own bound; in decode, a step that carries
one token per rank compiles for a capacity of one. Without the uniformity
fact, one variant at the largest bound remains.

\paragraph{Storage planning and dead communication.}
Storage planning sizes each region, through a dataflow pass, to the extent
that its reads and writes reach. A receive buffer that an expert
kernel reads through one record per source rank,
for example, needs one row per record instead of one per routed copy. On
Qwen3-235B, storage planning as a whole adds 45\% KV-cache capacity
(\S\ref{sec:Eval-MoE}).
An input that every reader masks by the same count may be bound to the
framework's own tensor instead of being copied, and an opaque output may be
written directly into the donated buffer that a later copy would fill.
Transfers in a generic record interface
that no consumer reads, such as returning metadata the combine never uses, are
removed with their storage.

\paragraph{Endpoint fan-out.}
A chain larger than one queue serves well is split across $k$ endpoints of the
same peer, with $k$ taken from the chain's byte bound and a per-platform
profile measured once. Each lane sends its own row slice and signal, and
receive coalescing joins the receiver's waits for every lane in one program
per peer. An unknown platform, or a decode-sized bound, gets one lane.
Profiles are measured for B200 GPUs with ConnectX-7 NICs; our H100 platform
has none, so its chains use one lane.

\subsection{Queue IR and Backend Details}
\label{sec:App-Backends}

\paragraph{Queue IR attributes.}
\code{submit} records the passes' plans as attributes: whether the
completion is deferred, whether one thread owns the queue, whether a block
barrier must order the payload's producers before it, and whether the
submission also announces commands built by other threads
(Listing~\ref{lst:queue-ir}). Verifiers check that each submission stays on
one endpoint and preserves the token order.

\paragraph{Issuer analysis.}
The analysis accepts an endpoint index that is a one-to-one function of the
program ID along one grid axis, possibly shifted, scaled, or rotated as
$(\mathit{pid}+c) \bmod N$, or linearized over two axes as
$\mathit{lane} \cdot W + \mathit{peer}$. The proof holds only for the grid
extents it assumes, e.g., a rotation is a permutation only when the grid's
extent is $N$, so the kernel asserts them at run time.

\begin{table}[t]
  \centering
  \small
  \caption{How each backend realizes submission, strong publication, and
  completion.}
  \label{tab:backends}
  \setlength{\tabcolsep}{3pt}
  \begin{tabular}{@{}l>{\raggedright\arraybackslash}p{0.255\columnwidth}>{\raggedright\arraybackslash}p{0.245\columnwidth}>{\raggedright\arraybackslash}p{0.22\columnwidth}@{}}
    \toprule
    \textbf{Backend} & \textbf{Submission} & \textbf{Strong signal} &
    \textbf{Completion} \\
    \midrule
    mlx5 & GPU writes WQEs; GPU or host rings & Fenced write or remote
    atomic & One CQE per planned interval \\
    EFA & GPU writes WQEs to the NIC or a ring the host copies &
    Retire payload, then write &
    Count CQEs \\
    Proxy & GPU writes commands to a host ring & Host-posted, per provider &
    Host-published sequence \\
    Peer memory & Peer loads and stores & Release exchange & None \\
    \bottomrule
  \end{tabular}
\end{table}

\paragraph{Host proxy.}
For NICs and platforms without GPU-side queues, the GPU writes compact
32-byte commands into a host-visible ring and publishes each batch with one
release store. The command holds source, destination, length, keys, and opcode.
A host thread posts the batch with one verbs or libfabric
call and publishes a completion sequence that \code{poll} reads on the GPU.

\paragraph{Peer memory.}
Payloads of at least 256 bytes use a cooperative copy: a Triton tensor of
8-byte elements that layout assignment distributes over all warps of the
program, copied in chunks of up to 16\,KiB that shrink to fit a payload of
constant size. Smaller or misaligned payloads use a per-thread loop, and
detached Tier-1 functions use a one-warp variant. The copy is complete when
its stores are issued, so \op{wait} and \op{drain} are no-ops.

\subsection{Doorbell Forwarding}
\label{sec:App-Doorbells}
With CPU-assisted submission, the GPU still builds WQEs in its own memory
and only delegates the doorbell store to a host thread.
By default, every queue a process creates through DEVX shares one doorbell
register and a lock for threads that ring it. This reduces the
rate at which doorbells are forwarded, as host threads serialize on
the register.
Giving each endpoint its own doorbell page instead and replacing the lock
with atomic writes allows threads to reach the NIC in parallel. With four
submitting threads per GPU on 128 GPUs, this reduces the all-to-all round
at 256\,B per peer 1.8$\times$ from 45.0 to 24.5\,\si{\micro\second}.
From 128\,KiB per peer, it makes the round 4--7\% faster, but slows down
6--14\% at 16--64\,KiB. This configuration is therefore not selected
by default. We note that at 1--4\,MiB per peer on 32 GPUs,
DEVX-created queues, which NVSHMEM's IBGDA and NCCL's GDAKI also use,
trail queues the kernel driver creates by 3--4\%.

\begin{table*}[t]
  \centering
  \footnotesize
  \caption{Evaluation platforms and software versions.}
  \label{tab:platforms}
  \setlength{\tabcolsep}{3.5pt}
  \begin{tabular}{@{}l>{\raggedright\arraybackslash}p{0.27\textwidth}>{\raggedright\arraybackslash}p{0.27\textwidth}>{\raggedright\arraybackslash}p{0.3\textwidth}@{}}
    \toprule
    & \textbf{H100} (\S\ref{sec:Eval-Libraries}--\S\ref{sec:Eval-DSL},
    App.~\ref{sec:App-Eval}, \ref{sec:App-DSL}) & \textbf{B200}
    (\S\ref{sec:Eval-Libraries}, \S\ref{sec:Eval-MoE},
    App.~\ref{sec:App-Eval}) &
    \textbf{MI355X} (\S\ref{sec:Eval-Ablation}, App.~\ref{sec:App-AMD}) \\
    \midrule
    GPUs & 4 H100 per node, 1--64 nodes & 8 B200 per node, 1--4 nodes &
    8 MI355X per node, 2 nodes \\
    Intra-node & NVLink & NVLink & xGMI \\
    Network & ConnectX (mlx5), 200\,Gb/s InfiniBand & ConnectX-7 (mlx5),
    400\,Gb/s InfiniBand & AMD Pollara 400, 400\,Gb/s RoCEv2 \\
    Submission & CPU-assisted &
    GPU-direct & Host proxy or GPU-direct (network), peer memory (xGMI) \\
    Driver, runtime & Driver 595.91.07, CUDA 12.6 & Driver 610.57.04, CUDA
    13.0 & ROCm 7.14.0, HIP 7.14 \\
    Triton & 3.8.0 & 3.8.0 & 3.8.0 \\
    Baselines & NCCL 2.31.2, NVSHMEM 3.4.5; DSLs: NVSHMEM4Py 0.4.1 over
    NVSHMEM 3.8.0, Triton-distributed with NVSHMEM 3.6.5 & NCCL 2.32.3,
    DeepEP 2.5.0, vLLM with the \toolname adapter & rocSHMEM 3.5.0,
    RCCL 2.30.4 \\
    \bottomrule
  \end{tabular}
\end{table*}
\section{Platforms}
\label{sec:App-Platforms}
Table~\ref{tab:platforms} lists the hardware, submission path, and software
versions of the three platforms and the sections that use each.

\section{Preliminary AMD MI355X Results}
\label{sec:App-AMD}

\begin{figure}[t]
  \centering
  \includegraphics[width=\columnwidth]{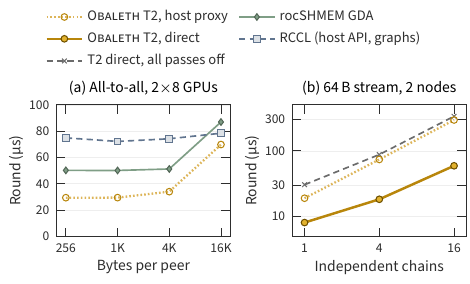}
  \caption{\toolname on AMD MI355X GPUs in two nodes, one NIC each. (a)
    All-to-all with host-proxy submission. (b) A stream of
    independent 64\,B chains whose buffers are reused every 16 rounds.
  Medians of five interleaved runs.}
  \label{fig:mi355x}
\end{figure}

We ported \toolname to AMD MI355X GPUs with AMD Pollara RoCE NICs: Tier-2
kernels compile to AMDGCN through Triton's AMD backend, and Tier-1 is
generated as a HIP device library. Through the host
proxy (Table~\ref{tab:backends}), \toolname has the lowest round trip up to
32\,KiB, both across nodes (28.3 against 36.6\,\si{\micro\second} for
rocSHMEM at 8\,B) and over xGMI. At 256\,B per peer, its all-to-all across
two nodes is 1.24--1.70$\times$ faster than rocSHMEM's and
2.2--2.5$\times$ faster than RCCL's on 8 and 16 GPUs
(Figure~\ref{fig:mi355x}a). As the host thread builds the work requests,
few passes apply to the proxy, and we implemented no passes specific to AMD.
Figure~\ref{fig:mi355x}b shows the GPU-direct backend, in which the GPU
submits to the NIC itself, with and without the passes
(\S\ref{sec:Eval-AMD}).

\section{Additional Results}
\label{sec:App-Eval}

\begin{figure}[t]
  \centering
  \includegraphics[width=\columnwidth]{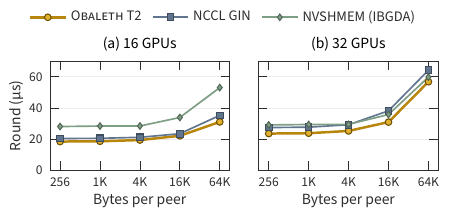}
  \caption{All-to-all on B200 GPUs with GPU doorbells, 8 GPUs per node. One
    configuration per library (NCCL GIN with warp-issued fused puts;
  NVSHMEM's collective). Medians of three runs.}
  \label{fig:app-b200}
\end{figure}

\paragraph{GPU doorbells.}
On B200, Tier-2 leads both libraries at every size up to 64\,KiB per peer
(Figure~\ref{fig:app-b200}): by 1.06--1.23$\times$ over NCCL GIN and
1.05--1.71$\times$ over NVSHMEM.

\paragraph{Streaming.}
Without completions, NVSHMEM posts from 128--256 threads per queue and
streams 257\,M\,msg/s on 16 queues, the NIC's limit, against
213\,M\,msg/s for Tier-2, which writes every WQE of a batch from one thread
(about 74\,ns per put). On 64 queues, Tier-2 reaches the limit as well
(256\,M\,msg/s).

\section{Additional Kernel DSL Results}
\label{sec:App-DSL}

Figures~\ref{fig:app-dsl} and~\ref{fig:app-dsl-moe} give the absolute values
behind Figures~\ref{fig:libraries}h and~\ref{fig:dsl-moe}, with every
configuration of the MoE comparison of \S\ref{sec:Eval-DSL}.

\paragraph{Setup.}
The DSL kernels match the CUDA versions call for call. CuTe DSL kernels call
Tier-1 as an external device library, or NVSHMEM through its CuTe bindings
(NVSHMEM4Py). Triton kernels link Tier-1 as device bitcode or are Tier-2
kernels, and the baseline is Triton-distributed, which links NVSHMEM's
bitcode into Triton kernels. Each library runs the best configuration it
ships with, including warp-issued and fused put-with-signal operations and
NVSHMEM's collective where its bindings expose it.

\paragraph{Microbenchmarks.}
NVSHMEM's CuTe bindings take 16.0\,\si{\micro\second} per round trip,
3.0\,\si{\micro\second} more than NVSHMEM from CUDA in the same runs.
Tier-1's message rate is below NVSHMEM's at 16 CTAs, as it rings a doorbell
per put, and 2.2--3.6$\times$ above it at 64 CTAs. Triton-distributed takes
its faster transport per benchmark: IBRC for the round trip and all-to-all,
IBGDA for the message rate. At 4\,MiB per peer on 8 GPUs, both tiers reach
41\,GB/s from both languages, against 11--19\,GB/s for NVSHMEM; the drop is
NVSHMEM's own, as its per-peer kernels and its 3.8.0 collective drop the same
way from CUDA. \toolname
allocates 4\,MiB of device memory per rank in every language, against 2126
and 1614\,MiB for NVSHMEM4Py and Triton-distributed.

\paragraph{MoE dispatch and combine.}
Both models are top-8 of 128 experts (hidden sizes 4096 and 2048) with FP8
in 128-element scale groups. Every configuration receives the same seeded
routing, quantizes to FP8 inside its dispatch, and must return the same
combined output. Besides \toolname's compiled kernels (\S\ref{sec:Eval-MoE})
and DeepEP's \code{ElasticBuffer} (version~2, over NCCL's device API), we run
Triton-distributed's low-latency EP layer and cuTile kernels that pack and
reduce around either Tier-2's send kernel or NCCL's \code{all_to_all}. From
64 tokens per rank on Qwen3-235B, DeepEP's dispatch is nearly flat, while
\toolname's NVLink send programs spill registers from eight tokens. In cuTile,
Tier-2 beats NCCL by 1.05--1.87$\times$ in dispatch from eight tokens per
rank and by 1.11--2.43$\times$ in combine at every size. From 32
tokens, the combine also beats the compiled one by up to 1.5$\times$,
so the cost at large batches lies in the compiled kernels, not the
transport. Triton-distributed over IBGDA (median of three runs) is
1.6--6.6$\times$ slower than \toolname's compiled kernels in dispatch
and 2.0--11$\times$ in combine.

\begin{figure*}[t]
  \centering
  \includegraphics[width=\textwidth]{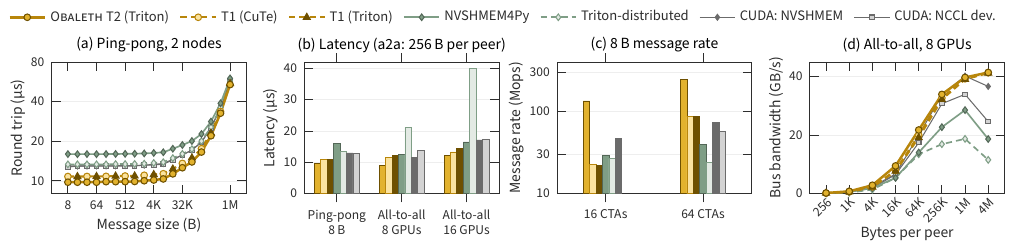}
  \caption{\toolname called from CuTe DSL and Triton kernels, against NVSHMEM
    as each language ships it (NVSHMEM4Py; Triton-distributed) and, in gray,
  CUDA libraries. Tier-1 uses a GPU-scope doorbell.}
  \label{fig:app-dsl}
\end{figure*}

\begin{figure*}[t]
  \centering
  \includegraphics[width=\textwidth]{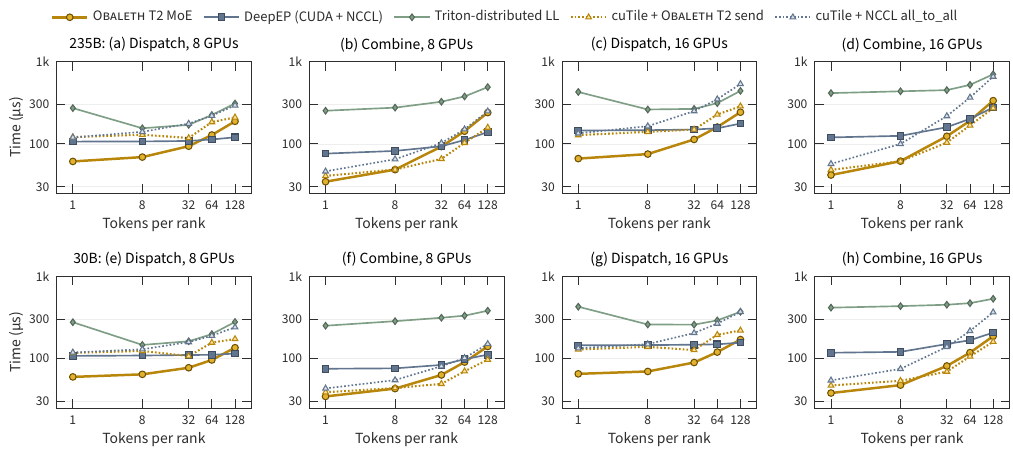}
  \caption{Dispatch and combine for one MoE layer on H100, Qwen3-235B-A22B
    (top) and Qwen3-30B-A3B (bottom). Median per-call time at the slowest
    rank, then the median of five interleaved runs. cuTile variants pack and
    reduce in cuTile
    kernels around \toolname's Tier-2 send kernel or NCCL's
    \code{all_to_all}. Triton-distributed runs over NVSHMEM's IBGDA
  transport (median of three runs).}
  \label{fig:app-dsl-moe}
\end{figure*}

\fi

\ifabappendix\else
\bibliography{references}
\bibliographystyle{mlsys2025}
\fi

\ifabappendix\else
\section{AI Usage Disclosure}
\label{sec:AI-Disclosure}

The authors used coding assistants with Claude and GPT models to
implement the compiler and associated parts described in this work,
and to design its experiments and plot the results. The same models
were used for light editing of the manuscript, i.e., typo fixes and
grammar polish, styling of listings and tables, ensuring compliance
with the venue guidelines.

\fi

\ifabpaper\else\ifabappendix\else
\clearpage
\toolnamewritecounters

\fi\fi

\end{document}